\documentclass[epj,nopacs]{svjour}
\usepackage{graphics}
\usepackage{enumitem}
\usepackage{dcolumn} 
\usepackage{bm} 
\usepackage{amssymb}
\usepackage{xcolor}
\usepackage[normalem]{ulem}
\usepackage{cancel}
\usepackage{amsmath}
\usepackage[english]{babel}
\newcommand{\be}{\begin{equation}}
\newcommand{\ee}{\end{equation}}

\newcommand{\de}{\mbox{d}}

\newcommand{\spl}{\be\begin{split}}
\newcommand{\ph}{\phantom{a}}
\newcommand{\pa}{\partial}

\begin{document}
\title{Local conformal symmetry in non-Riemannian geometry and the origin of physical scales}
\author{Marco de Cesare\inst{1}\thanks{marco.de\_cesare@kcl.ac.uk} \and John W. Moffat\inst{2}\thanks{jmoffat@perimeterinstitute.ca} \and Mairi Sakellariadou\inst{1,2}\thanks{mairi.sakellariadou@kcl.ac.uk}
}                     
\institute{Theoretical Particle Physics and Cosmology Group, Department of Physics, King's College London, University of London, Strand, London, WC2R 2LS, U.K. \and Perimeter Institute for Theoretical Physics, Waterloo, Ontario N2L 2Y5, Canada}
\date{Received: date / Revised version: date}
\abstract{
We introduce an extension of the Standard Model and General Relativity built upon the principle of local conformal invariance, which represents a generalization of a previous work by Bars, Steinhardt and Turok.
This is naturally realized by adopting as a geometric framework a particular class of non-Riemannian geometries, first studied by Weyl. The gravitational sector is enriched by a scalar and a vector field. The latter has a geometric origin and represents the novel feature of our approach. We argue that physical scales could emerge from a theory with no dimensionful parameters, as a result of the spontaneous breakdown of conformal and electroweak symmetries. We study the dynamics of matter fields in this modified gravity theory and show that test particles follow geodesics of the Levi-Civita connection, thus resolving an old criticism raised by Einstein against Weyl's original proposal.
\PACS{     04.50.-h   \and 11.25.Hf} 
} 
\maketitle
\section{Introduction}
The classical action of the Standard Model (SM) of particle physics is close to being conformally invariant. The only dimensionful coupling constants it features are given by the Higgs mass and its vacuum expectation value (vev), the latter setting the scale of electroweak (EW) symmetry breaking at $v\sim$ 246~GeV. Such a value is remarkably small compared to the Planck mass $M_{\rm P}\sim10^{19}$~GeV, which is set by the strength of the gravitational coupling. The huge gap between the two scales defines the hierarchy problem. A fourth dimensionful parameter, the cosmological constant, is responsible for the observed late acceleration of the Universe. The cosmological constant scale is $10^{-123}$ smaller than the Planck scale, leading to a second hierarchy problem in the SM coupled to gravity.

In this work, we embed the SM and General Relativity (GR) in a larger theory which exhibits local scale invariance classically. All couplings are therefore dimensionless. A mass scale arises through gauge fixing the conformal symmetry, from which all dimensionful couplings can be derived. Thus, all couplings which characterize fundamental physics at low energy scales are shown to have a common origin, in the same spirit as in Ref.~\cite{Bars:2013yba}. The role of EW symmetry breaking is crucial in this respect and is realized by means of a potential having the same form as the Higgs-dilaton potential, which was considered in Refs.~\cite{Bars:2013yba,Shaposhnikov:2008xb}.

A natural framework in which scale invariance can be realized as a local symmetry is given by a generalization of Riemannian geometry, known as Weyl geometry. A Weyl manifold is defined as an equivalence class of conformally equivalent Riemannian manifolds, equipped with a notion of parallel transport which preserves the metric only up to local rescalings~\cite{calderbank1997einstein}. Such non-Riemannian structures were first introduced by Weyl in pursuit of a unification of gravity and electromagnetism \cite{weyl:1918}. They were later reconsidered in an early paper by Smolin~\cite{Smolin:1979uz} in an attempt to reformulate gravity as a renormalizable quantum field theory. In this paper, as in Ref.~\cite{Smolin:1979uz}, Weyl geometry and conformal invariance are used to motivate the occurrence of new degrees of freedom in the gravitational sector and as guiding principles to build the action functional. Weyl geometry was later rediscovered independently by Cheng \cite{Cheng:1988zx}, who used it to formulate a model with no Higgs particle.

Conformal invariance imposes strong constraints on the terms that can appear in the action and enriches the gravitational sector with a scalar and a vector field. The theory thus obtained is a generalization of Brans-Dicke theory and of conformally invariant gravity theories, such as the one considered in Ref.~\cite{Bars:2013yba}. When the Weyl vector is pure gauge, the theory is equivalent to Brans-Dicke, of which it provides a geometric interpretation. This particular case has appeared in the literature under the name of Weyl Integrable Space-Time (WIST)~\cite{Romero:2012hs,Almeida:2013dba,Salim:1996ei}. However, in those works an additional assumption motivated by Ref.~\cite{Ehlers2012} is made about the free fall of test bodies, which marks a difference with Brans-Dicke. For applications of WIST to cosmology and to the study of spacetime singularities, see \emph{e.g.} Refs.~\cite{Lobo:2015zaa,Gannouji:2015vva}. Generalised scale invariant gravity theories were also obtained in Ref.~\cite{Padilla:2013jza}, by gauging the global conformal symmetry of (a subset of) the Horndeski action with the introduction of the Weyl vector.

Our framework is distinct from conformal gravity~\cite{Mannheim:1988dj,Maldacena:2011mk,Mannheim:2011ds}, where the affine connection is the Levi-Civita one also in the gravity sector. In that case, conformal symmetry is implemented by taking the square of the Weyl tensor as the Lagrangian. The Weyl tensor squared also appears in the bosonic spectral action in the context of noncommutative geometry~\cite{Kurkov:2014twa,Sakellariadou:2016dfl} and in the computation of the (formal) functional integral for quantum gravity~\cite{Hooft:2010ac}. 

In this paper we construct an effective field theory with local conformal invariance and show how the SM of particle physics and GR are recovered from it by means of a two-stage spontaneous symmetry breaking. Our proposal is based on a generalisation of Riemannian geometry, namely Weyl geometry, which leads to the introduction of new gravitational degrees of freedom: a scalar field $\phi$ and the Weyl vector $B_{\mu}$. There has been a recent surge of interest in the role of conformal symmetry in gravitational physics, see \emph{e.g.} Refs.~\cite{Bars:2013yba,Hooft:2010ac,Gielen:2015uaa,Gielen:2016fdb}, suggesting that it may play a role in Quantum Gravity. It is therefore possible that the gravitational theory emerging in the classical limit would also display such a symmetry. In this sense, our work is motivated by similar considerations to the ones usually put forward for the introduction of modified gravity theories, see \emph{e.g.} Refs.~\cite{Sotiriou:2007yd,Sotiriou:2008rp,Capozziello:2011et}. In addition, we adopt local conformal invariance as a guiding principle in selecting the action functional \emph{and} the geometric structure of spacetime. The enriched gravitational sector is to be interpreted as purely classical. SM fields are quantized as usual on the classical curved background defined by $g_{\mu\nu}$ \emph{and} $\phi$, $B_\mu$. This can be considered as a generalization of what is usually done in conventional quantum field theory on curved spacetimes.

We would like to mention that the same geometric setting and symmetry breaking process were considered in an unpublished work by Nishino and Rajpoot\footnote{Courtesy of the authors.} \cite{Nishino:2004kb}, although their motivations were different. In that paper the authors point out issues with renormalisability and unitarity in their model. Other aspects of the quantum theory are discussed in Refs.~\cite{Nishino:2009in,Nishino:2011zz}. Furthermore, the authors of Ref.~\cite{Nishino:2004kb} claim that local conformal invariance ``inevitably leads to the introduction of General Relativity". We disagree with their statement. Local conformal invariance of the SM sector only leads to the introduction of the Weyl vector, which is also not enough to determine the affine connection of a Weyl spacetime. Moreover, in our approach there are no issues with renormalisability and unitarity since our model is a \emph{classical} effective field theory.

The plan of the paper is the following. In Section~\ref{sec:Weyl} we recall the fundamentals of Weyl geometry and introduce the notation. In Section~\ref{Theory} we formulate our effective field theory and discuss how the Higgs and the scalar fields couple to gravity. In Section~\ref{EW SSB} we discuss the EW symmetry breaking and show how the dimensionful couplings which govern low energy physics are determined from the parameters of the model and from the scale of ``broken'' conformal symmetry. In Section~\ref{Sec:CouplingSM} we study the other sectors of the SM and show that no further modification is needed to achieve compatibility with local conformal invariance. In Section~\ref{Sec:Fluids} we consider the approximate description of matter as a fluid, following from the underlying field theory of Section~\ref{Sec:Fluids}, and use it to derive the equations of motion of test bodies. In Section~\ref{Section:Alternative} we consider an alternative, phenomenological model for the motion of macroscopic test bodies.
We review our results in the Conclusion, Section~\ref{Conclusions}, where we also examine the relation between our proposal and earlier ones in the literature.
In Section~\ref{Sec:Discussion} we discuss important features of our results and point at directions for future work.
\section{\label{sec:Weyl} Weyl geometry}
We follow Ref.~\cite{Smolin:1979uz} to introduce the basic concepts and notation, although our conventions for the Riemann tensor are different and coincide with those in Ref.~\cite{Wald:1984rg}. A Weyl manifold is a conformal manifold, equipped with a torsionless connection, called Weyl connection, that preserves the conformal structure. We thus consider a torsion-free affine connection which satisfies the condition
\be\label{eq:DefWeylConnection}
\nabla_{\lambda}g_{\mu\nu}=B_{\lambda}\;g_{\mu\nu}\; .
\ee
Equation~(\ref{eq:DefWeylConnection}) defines the Weyl connection $\nabla_{\lambda}$, which is a particular case of a connection with non-metricity (see \emph{e.g.} Ref.~\cite{Sotiriou:2006qn}). The Levi-Civita connection will instead be denoted by $D_{\lambda}$.
The connection coefficients are given by
\be\label{eq:ConnectionCoefficients}
\Gamma^{\sigma}_{\mu\nu}=\left\{ {\sigma \atop \mu\;\nu} \right\}-\frac{1}{2}\left(\delta^{\sigma}_{\mu}\,B_{\nu}+\delta^{\sigma}_{\nu}\,B_{\mu}-g_{\mu\nu}\,B^{\sigma}\right)\;.
\ee
Under a local conformal transformation\footnote{Local conformal transformations are also known as Weyl rescalings.}
\be\label{eq:LocalConformalMetric}
g_{\mu\nu}\rightarrow\tilde{g}_{\mu\nu}=\Omega^2g_{\mu\nu}~,
\ee
the Weyl one-form $B_\mu$ transforms as an Abelian gauge field
\be\label{eq:TransformationLawWeylVector}
B_\mu\rightarrow\tilde{B}_\mu=B_\mu+2\Omega^{-1}\nabla_{\mu}\Omega~,
\ee
so that the condition given by Eq.~(\ref{eq:DefWeylConnection}) is preserved. The connection coefficients in Eq.~(\ref{eq:ConnectionCoefficients}) are by definition conformally invariant.

The components of the Riemann curvature tensor in a local chart are given by
\be\label{eq:defineRiemann}
R_{\mu\nu\rho}^{\ph\ph\ph\sigma}=-\pa_\mu\Gamma^{\sigma}_{\nu\rho}+\pa_\nu\Gamma^{\sigma}_{\mu\rho}-\Gamma^{\sigma}_{\mu\kappa}\Gamma^{\kappa}_{\nu\rho}+\Gamma^{\sigma}_{\nu\kappa}\Gamma^{\kappa}_{\mu\rho}\;.
\ee
The Riemann tensor satisfies the following properties, as in the standard case:
\begin{enumerate}[label=\alph*)]
\item ~$R_{\mu\nu\rho}^{\ph\ph\ph\sigma}=-R_{\nu\mu\rho}^{\ph\ph\ph\sigma}$~;
\item ~$R_{[\mu\nu\rho]}^{\ph\ph\ph\ph\sigma}=0$, which follows from the symmetry of the connection coefficients, \emph{i.e.} the vanishing of the torsion~;
\item ~$\nabla_{[\lambda}R_{\mu\nu]\rho}^{\ph\ph\ph\ph\sigma}=0$~.
\end{enumerate}
Antisymmetry over the last two indices, which holds in the standard case, is replaced by 
\be\label{eq:FourhtPropertyRiemann}
R_{\mu\nu\rho\sigma}=-R_{\mu\nu\sigma\rho}+H_{\mu\nu}\;g_{\rho\sigma}\;,
\ee
where $H_{\mu\nu}$ is the field strength of $B_{\mu}$, defined as in electromagnetism
\be\label{eq:FieldStrength}
H_{\mu\nu}=\nabla_{\mu}B_{\nu}-\nabla_{\nu}B_{\mu}=
\pa_{\mu}B_{\nu}-\pa_{\nu}B_{\mu}\; .
\ee
The Riemann curvature of the Weyl connection, defined by Eq.~(\ref{eq:DefWeylConnection}), has the following expression\footnote{Square brackets denote antisymmetrization, as in $T_{[\mu\nu]}=\frac{1}{2}\left(T_{\mu\nu}-T_{\nu\mu}\right)$.}
\be\label{eq:RiemannDefinition}
\begin{split}
R_{\mu\nu\rho}^{\ph\ph\ph\sigma}=&R_{\mu\nu\rho}^{0\ph\ph\sigma}+\delta^\sigma_{[\nu} D_{\mu]} B_\rho+\delta^\sigma_\rho D_{[\mu}B_{\nu]}-g_{\rho[\nu}D_{\mu]}B^{\sigma}\\ &-\frac{1}{2}\left(B_{[\mu}\,g_{\nu]\rho}B^{\sigma}+\delta^{\sigma}_{[\mu}\,B_{\nu]}B_{\rho}+g_{\rho[\mu}\,\delta^{\sigma}_{\nu]}B_\lambda B^\lambda \right)~.
\end{split}
\ee
In the last equation, $R_{\mu\nu\rho}^{0\ph\ph\sigma}$ is the Riemann tensor of the Levi-Civita connection. It can be computed from Eq.~(\ref{eq:defineRiemann}), using the Christoffel symbols as the connection coefficients
\be\label{eq:defineOrdinaryRiemann}
\begin{split}
R_{\mu\nu\rho}^{0\ph\ph\sigma}=-\pa_\mu\left\{ {\sigma \atop \nu\;\rho} \right\}+\pa_\nu\left\{ {\sigma \atop \mu\;\rho} \right\}-\left\{ {\sigma \atop \mu\;\kappa} \right\}\left\{ {\kappa \atop \nu\;\rho} \right\}\\+\left\{ {\sigma \atop \nu\;\kappa} \right\}\left\{ {\kappa \atop \mu\;\rho} \right\}\;.
\end{split}
\ee
Defining the Ricci tensor by contracting the second and the fourth indices of the Riemann curvature in Eq.~(\ref{eq:RiemannDefinition})
\be
R_{\mu\nu}=R_{\mu\sigma\nu}^{\ph\ph\ph\sigma}\; ,
\ee
one has
\be\label{eq:RicciTensorExpand}
\begin{split}
R_{\mu\nu}=R^0_{\mu\nu}+D_\mu B_\nu +\frac{1}{2}H_{\mu\nu}+\frac{1}{2}g_{\mu\nu}D_{\sigma}B^{\sigma}\\
+\frac{1}{2}\left(B_\mu B_\nu-g_{\mu\nu}B_\sigma B^\sigma\right)~.
\end{split}
\ee
Note that, as a consequence of Eq.~(\ref{eq:FourhtPropertyRiemann}), the Ricci tensor is not symmetric. In fact, one has
\be
R_{[\mu\nu]}=H_{\mu\nu}\; .
\ee
The Riemann and the Ricci tensors are by definition conformally invariant.
The Ricci scalar is then defined as
\be\label{eq:RicciScalarDefine}
R=g^{\mu\nu}R_{\mu\nu}~.
\ee
Under a conformal transformation the Ricci scalar reads
\be\label{eq:RescalingScalarCurvature}
R\rightarrow\tilde{R}=\Omega^{-2}R~.
\ee
Substituting  Eq.~(\ref{eq:RicciTensorExpand}) into Eq.~(\ref{eq:RicciScalarDefine}), the Ricci scalar is
\be\label{eq:RicciScalarExpand}
R=R^0+3D_\mu B^\mu-\frac{3}{2}B_\mu B^\mu~,
\ee
where $R^0$ is the Ricci scalar computed from the ordinary Riemann curvature, Eq.~(\ref{eq:defineOrdinaryRiemann}).

\section{A geometric scalar-vector-tensor theory}\label{Theory}
\subsection{The simplest model}\label{sec:SimpleModel}
Our aim is to build an action functional for gravity which is conformally invariant. We will follow Smolin for its derivation~\cite{Smolin:1979uz}. From Eqs.~(\ref{eq:LocalConformalMetric}),~(\ref{eq:RescalingScalarCurvature}) we see that the simplest action displaying such property is
\be\label{eq:Action}
S_{\rm g}=\int \de^4 x\sqrt{-g}\; \xi_\phi\phi^2 R~,
\ee
where $\xi_\phi$ is a coupling constant and $\phi$ is a real scalar field transforming under local rescalings, Eq.~(\ref{eq:LocalConformalMetric}), according to its canonical dimensions\footnote{Note that the transformation properties of the volume element $\de^4 x\sqrt{-g}$ under conformal transformations are determined by those of the determinant of the metric (coordinates are not rescaled). From Eq.~(\ref{eq:LocalConformalMetric}) we have $\sqrt{-g}\to\Omega^4\sqrt{-g}$. This is important when checking conformal invariance of the action~(\ref{eq:Action}).}
\be
\phi\rightarrow\tilde{\phi}=\Omega^{-1}\phi~.
\ee
We impose the further requirements that the equations of motion shall contain no derivatives higher than second order and no inverse powers of the scalar field $\phi$ shall appear in the action.  Equation~(\ref{eq:Action}) is therefore singled out as the unique action satisfying the above conditions, in the case of a single non-minimally coupled real scalar field. The scalar field contributes another term to the action
\be\label{eq:PhiSector}
S_{\rm s}=\int \de^4 x\sqrt{-g}\; \left[-\frac{\omega}{2}\; g^{\mu\nu}\left(\pa_\mu\phi+\frac{1}{2}B_{\mu}\phi\right) \left(\pa_\nu\phi+\frac{1}{2}B_{\nu}\phi\right)\right]~,
\ee
where a minimal coupling to the Weyl one-form $B_{\mu}$ has been considered in order to make the action consistent with the principle of local conformal invariance, and $\omega$ is the Brans-Dicke parameter. Lastly, $B_{\mu}$ is made dynamical by adding a kinetic term to the action
\be\label{eq:YMaction}
S_{\rm v}=\int \de^4 x\sqrt{-g}\; \left[-\frac{1}{4f^2}\; H_{\mu\nu}H^{\mu\nu}\right]~,
\ee
in complete analogy with electrodynamics. The field strength $H_{\mu\nu}$  of $B_\mu$ is defined as in Eq.~(\ref{eq:FieldStrength}). The action (\ref{eq:YMaction}) is the Yang-Mills action for an Abelian gauge field. It represents the most natural choice which is compatible with local scale invariance, since the Yang-Mills action is conformally invariant in four dimensions. The parameter $f$ is a universal coupling costant. The action $S_{\rm g}+S_{\rm s}+S_{\rm v}$ defines the extended gravitational sector of the theory.

The scalar field $\phi$ introduced above can be interpreted as a dilaton. In fact, it gives the strength of the gravitational coupling. However, since we are considering \emph{local} conformal symmetry, the dilaton $\phi$ can be eliminated by an appropriate gauge fixing, as we will show in the next section. Gauge fixing also yields a massive vector $B_\mu$ in the spectrum, thus preserving the total number of degrees of freedom. We should point out that there are other gauge choices in which $\phi$ is instead dynamical, such as those considered in Ref.~\cite{Bars:2015trh}.
\subsection{Coupling the Higgs field to gravity}
The theory given in the previous section can be immediately extended to include the Standard Model Higgs field. In fact, we will show that it is possible to embed the SM in a theory with local conformal invariance. As a result, all dimensionful parameters such as the gravitational constant, the Higgs vev, the Higgs mass, and the cosmological constant will all have a common origin. The tensor sector is given by
\be\label{eq:PhiHiggsTensorSector}
S_{\rm g}=\int \de^4 x\sqrt{-g}\; \left(\xi_{\phi}\;\phi^2+2\xi_H\; H^{\dagger}H\right) R~,
\ee
where $\xi_{\phi}$, $\xi_H$ are dimensionless couplings. The Higgs kinetic term, including a minimal coupling to the Weyl one-form, is given by
\be\label{eq:HiggsSector}
\begin{split}
&S_{\rm H}=\\
&\int \de^4 x\sqrt{-g}\; \left[-g^{\mu\nu}\left(\pa_\mu H^{\dagger}+ \frac{1}{2}B_{\mu} H^{\dagger}\right)\left(\pa_\nu H+ \frac{1}{2}B_{\nu} H\right)\right]~.
\end{split}
\ee
When introducing Yang-Mills connections corresponding to the SM gauge group, partial and covariant derivatives are replaced by gauge covariant derivatives.
 
We can then introduce a Higgs-dilaton potential as in Ref.~\cite{Shaposhnikov:2008xb},
\be\label{eq:HiggsDilatonPotential}
V(\phi,H)=\frac{\lambda}{4}\left(H^{\dagger}H-\kappa^2\phi^2\right)^2+\lambda^
{\prime}\phi^4~,
\ee
where $\lambda$, $\lambda^{\prime}$, $\kappa$ are dimensionless parameters. 

Fixing the gauge in such a way that $\phi$ takes a constant value $\phi_0$ everywhere in spacetime, the Higgs-dilaton potential takes the form of the usual Mexican hat potential, including a cosmological constant term, namely
\be\label{eq:Higgs-Dilaton}
V(\phi_0,H)=\frac{\lambda}{4}\left(H^{\dagger}H-\kappa^2\phi_0^2\right)^2+\lambda^{\prime}\phi_0^4~.
\ee
We can write the Higgs doublet in the unitary gauge
\be
H=\frac{1}{\sqrt{2}}\begin{pmatrix} 0\\ h\end{pmatrix}~.
\ee
It is then readily seen that EW symmetry breaking fixes the values of the gravitational coupling $G$, the Higgs vev $v$, as well as the Higgs mass $\mu$, and the cosmological constant $\Lambda$, in terms of the scale of conformal symmetry breaking $\phi_0$, as (cf. Ref.~\cite{Bars:2013yba})
\be
\begin{split}
\frac{\Lambda}{8\pi G}=\lambda^{\prime}\phi_0^4~&,~ \hspace{1em} \frac{v^2}{2}=\kappa^2\phi_0^2~,\\~ \hspace{1em} \frac{1}{16\pi G}=\xi_{\phi}\;\phi_0^2+\xi_H\; v^2~&,~ \hspace{1em} \mu^2=-\lambda\kappa^2\phi_0^2~.
\end{split}
\ee

The conformally invariant theory of gravity given here can be seen as a generalization of other theories with local conformal invariance proposed in the literature. Considering Eq.~(\ref{eq:RicciScalarExpand}), we can rewrite the total action given by the sum of the $S_{\rm g}$, $S_{\rm s}$, $S_{\rm H}$ and $S_{\rm v}$ contributions from Eqs.~(\ref{eq:PhiHiggsTensorSector}), (\ref{eq:PhiSector}), (\ref{eq:HiggsSector}) and (\ref{eq:YMaction}), respectively, and including the potential Eq.~(\ref{eq:HiggsDilatonPotential}) as
\be
\begin{split}
S=&\int \de^4 x\sqrt{-g}\; \bigg[\left(\xi_{\phi}\;\phi^2+2\xi_H\; H^{\dagger}H\right) R^0-\frac{\omega}{2}\pa^\mu\phi \pa_\mu\phi\\ &-\frac{1}{2}(\omega+12\xi_{\phi})\phi B^\mu\pa_\mu\phi -\frac{1}{8}(\omega+12\xi_\phi)\phi^2B_\mu B^\mu\\& -\pa^\mu H^{\dagger}\pa_\mu H-\frac{1}{2}(1+12\xi_H)B^\mu(H^{\dagger}\pa_\mu H+\pa_\mu H^{\dagger} H)\\&-\frac{1}{4}(1+12\xi_H)H^{\dagger}H\; B_\mu B^\mu-\frac{1}{4f^2}\; H_{\mu\nu}H^{\mu\nu}
\\&- \frac{\lambda}{4}\left(H^{\dagger}H-\kappa^2\phi^2\right)^2-\lambda^
{\prime}\phi^4\bigg]~,
\end{split}
\ee
up to a surface term.
\section{EW symmetry breaking and the scalar-tensor-vector\\gravity}\label{EW SSB}
As a consequence of the spontaneous breakdown of conformal and EW symmetries, the vector $B^{\mu}$ acquires a mass. This can be seen by looking at Eqs.~(\ref{eq:PhiSector}),~(\ref{eq:PhiHiggsTensorSector}),~(\ref{eq:HiggsSector}) and taking into account Eq.~(\ref{eq:RicciScalarExpand}). In fact, excluding interactions with other matter fields and with the Higgs boson, the action of $B^{\mu}$ reads
\be\label{eq:MassiveVectorAction}
S_{\rm v}=\int \de^4 x\sqrt{-g}\; \left[-\frac{1}{4f^2}\; H_{\mu\nu}H^{\mu\nu}-\frac{1}{2}m_B^2\; B_{\mu}B^{\mu}\right]~,
\ee
with
\be
m_B^2= 3\left(\xi_{\phi}\;\phi_0^2+\xi_H\; v^2\right)+\frac{\omega}{4}\phi_0^2+\frac{v^2}{4}=\frac{3}{16\pi G}+\frac{v^2}{4}\left(\frac{\omega}{2\kappa^2}+1\right)~.
\ee
It is possible to rewrite the action of the vector field in canonical form, by expanding the first term in Eq.~(\ref{eq:MassiveVectorAction}) and rescaling the field as $B^{\mu}\rightarrow f\; B^{\mu}$. We have
\be\label{eq:ProcaActionBField}
\begin{split}
S_{\rm v}= \int \de^4 x\sqrt{-g}\; \bigg[ -\frac{1}{2}\Big( (D^{\mu}B^{\nu})(D_{\mu}B_{\nu})&-(D^{\nu}B^{\mu})(D_{\mu}B_{\nu})\Big)\\ & -\frac{1}{2}f^2 m_B^2\; B_{\mu}B^{\mu}\bigg]~.
\end{split}
\ee
Hence, the physical mass squared of the vector is given by
\be
m_{\rm v}^2=f^2 m_B^2~.
\ee
Equation~(\ref{eq:ProcaActionBField}) is the Proca action in a curved spacetime. Sources $j^{\mu}$ for the field $B_{\mu}$ come from the other sectors of the theory; they are covariantly conserved, $D_{\mu}j^{\mu}=0$, as a consequence of the minimal coupling prescription. From the equations of motion one gets the subsidiary condition $D_{\mu}B^{\mu}=0$ (since $m_v^2\neq0$), which restricts the number of degrees of freedom of the vector field to three, namely two transverse modes and a longitudinal mode. Hence, counting degrees of freedom before and after the breaking of conformal invariance gives the same result. In analogy with the Higgs mechanism, we can say that the vector field $B^{\mu}$ acquires a mass and a longitudinal polarization mode as a result of conformal symmetry breaking. The dilaton $\phi$ can be completely decoupled from the theory by choosing a suitable gauge, as it happens for the Goldstone boson in the unitary gauge (see however the remark at the end of Section~\ref{sec:SimpleModel}). In fact, a stronger result holds: the kinetic term of $\phi$ is identically vanishing, which makes the field non-dynamical. Only its constant value $\phi_0$ appears in all equations written in this gauge. 

Before closing this section, we want to specify the connection between our model and the ones in the literature about conformal invariance in gravity and cosmology. Choosing the particular values of the parameters $\xi_H=\frac{\xi_\phi}{\omega}=-\frac{1}{12}$, the Higgs and the dilaton fields are completely decoupled from the vector field, which yields the action
\be\label{eq:BarsTurokAction}
\begin{split}
S=\int \de^4 x\sqrt{-g}\; \bigg[&-\left(\frac{\omega}{12}\phi^2+\frac{1}{6}\; H^{\dagger}H\right) R^0\\ &-\frac{\omega}{2}\pa^\mu\phi \pa_\mu\phi -\pa^\mu H^{\dagger}\pa_\mu H
-V(\phi,H)\bigg]~,
\end{split}
\ee
with $V(\phi,H)$ the Higgs-dilaton potential given from Eq.~(\ref{eq:HiggsDilatonPotential}).
Equation~(\ref{eq:BarsTurokAction}) is the action of two scalar fields with conformal coupling to curvature; it is the model considered in Ref.~\cite{Bars:2013yba}, for $\omega=-1$. Writing the Higgs field in the unitary gauge, the action Eq.~(\ref{eq:BarsTurokAction})  can be also seen as equivalent to the conformally invariant two-field model of Ref.~\cite{Kallosh:2013hoa} with $\mbox{SO(1,1)}$ symmetry.
\section{Coupling to SM fields}\label{Sec:CouplingSM}
So far, we have focused our attention on the gravitational sector of the theory, given by the fields $g_{\mu\nu}$, $B_\mu$ and $\phi$, and considered their couplings to the Higgs doublet. In this section we will focus on their couplings to SM fields and study whether the framework of Weyl geometry introduces any modifications to such sectors. We will discuss separately the cases of gauge bosons and spin-$1/2$ fermions (leptons and quarks).

Let us consider a gauge field $A_\mu^a$, where $a$ is an internal index labelling components in the Lie algebra of the gauge group. Its kinetic term is given by the square of its field strength\footnote{$g$ is the gauge coupling constant, $f^{abc}$ are the structure constants of the gauge group. In the Abelian case the second term in Eq.~(\ref{eq:GaugeFieldStrength}) vanishes.},
defined using the affine connection $\nabla_\mu$
\be\label{eq:GaugeFieldStrength}
F_{\mu\nu}^a=\nabla_\mu A^a_\nu - \nabla_\nu A^a_\mu + g f^{abc} A_\mu^bA_\nu^c.
\ee
It is well known that for all symmetric (\emph{i.e.} torsion-free) connections $\nabla_\mu$ the above can be rewritten as
\be
\begin{split}
F_{\mu\nu}^a=&D_\mu A^a_\nu - D_\nu A^a_\mu + g f^{abc} A_\mu^bA_\nu^c\\
=&\pa_\mu A^a_\nu - \pa_\nu A^a_\mu + g f^{abc} A_\mu^bA_\nu^c~.
\end{split}
\ee
In particular, this is true in the case when $\nabla_\mu$ is the Weyl connection.  Hence, there is no direct coupling between the Weyl vector and gauge bosons. The kinetic term of the gauge boson $A_\mu^a$ is given by the standard Yang-Mills action
\be
S_{\rm YM} =-\frac{1}{4}\int \de^4 x\sqrt{-g}\; F^a_{\mu\nu}F^{a\,\mu\nu}~,
\ee
which is conformally invariant in four dimensions. %
 The scalar field $\phi$ is real in our model, therefore it does not couple to ordinary gauge fields through the minimal coupling prescription. Although it is certainly possible to generalize the model to allow for non minimal couplings, they can potentially spoil conformal invariance or renormalizability of the SM (or both).

The description of the dynamics of fermions on curved spacetime requires the introduction of a tetrad and of a spin connection. The action of a massless Dirac spinor is given by (see \emph{e.g.} \cite{Parker:2009uva})
\be\label{eq:ActionDirac}
S_{\rm Dirac}=\int  \de^4 x\sqrt{-g}\; i \overline{\psi}\gamma^c e^{\mu}_c\left(\pa_\mu+\frac{1}{8}[\gamma^a,\gamma^b]\,e_a^{\;\nu}(D_\mu e_{b\,\nu})\right)\psi~.
\ee
Observe that Eq.~(\ref{eq:ActionDirac}) uses the Levi-Civita connection $D_\mu$. The reason for this choice will be clear from the following.
Latin indices are used for the Lorentzian frame defined pointwise by the tetrad $e^a_\mu$
\be
e^a_{\,\mu} e_{a\,\nu}=g_{\mu\nu},\hspace{1em} e^a_{\,\mu} e^{b\,\mu}=\eta^{ab}~.
\ee
$\eta_{ab}$ is the Minkowski metric ${\rm diag}$(-1,1,1,1). The gamma matrices in Eq.~(\ref{eq:ActionDirac}) are the flat ones $\{\gamma^a,\gamma^b\}=2\eta^{ab}$.
Under a conformal transformation, each field in Eq.~(\ref{eq:ActionDirac}) transforms according to its conformal weight
\be
\psi\rightarrow \tilde{\psi}=\Omega^{-3/2}\psi,\quad \overline{\psi}\rightarrow \tilde{\overline{\psi}}=\Omega^{-3/2}\overline{\psi},\quad e^a_{\,\mu}\rightarrow\tilde{e}^a_{\,\mu}=\Omega\, e^a_{\,\mu}~.
\ee
It is possible to check by explicit computation that, under such a transformation, all terms involving derivatives of the function $\Omega$ cancel in Eq.~(\ref{eq:ActionDirac}). Hence, the action of a Dirac fermion defined using the Levi-Civita connection is conformally invariant. The same conclusion  can also be reached by looking at the square of the Dirac operator defined by Eq.~(\ref{eq:ActionDirac}). In this way, one finds a generalization of the Klein-Gordon equation with a non-minimal coupling to curvature, which turns out to be conformally invariant \cite{Parker:2009uva,konno:1988}.
In Ref.~\cite{Cheng:1988zx} the action of a Dirac particle was defined by considering a generalization of Eq.~(\ref{eq:ActionDirac}) which makes both terms in the bracket separately conformally invariant, when acting on $\psi$. Namely, the Weyl connection is considered instead of the Levi-Civita connection and the coupling to the Weyl vector is included, with the appropriate coupling constant given by the conformal weight of the spinor
\be\label{Eq:ActionDirac2}
\int  \de^4 x\sqrt{-g}\; i \overline{\psi}\gamma^c e^{\mu}_c\left(\pa_\mu+\frac{3}{4}B_\mu+\frac{1}{8}[\gamma^a,\gamma^b]\,e_a^{\;\nu}(\nabla_\mu e_{b\,\nu})\right)\psi~.
\ee
However, it turns out that this action is equal to the one in Eq.~(\ref{eq:ActionDirac}), since the terms involving the Weyl vector cancel exactly. More details are given in the Appendix.

We conclude this section by stressing that the requirement of local conformal invariance does not introduce new direct couplings of the elementary matter fields (with the only exception of the Higgs) with the new fields $\phi$ and $B_\mu$. Their interactions with leptons, quarks and gauge bosons can only be mediated by the gravitational field $g_{\mu\nu}$ or the Higgs field. This has important implications for the dynamics of matter in a gravitational field. 

\section{Motion of fluids and test particles}\label{Sec:Fluids}
In the previous section we showed that the dynamics of free vector and spinor fields is determined solely by the Levi-Civita connection. The only field in the gravitational sector with whom they can interact directly is the metric tensor $g_{\mu\nu}$. A description of matter which is particularly convenient for applications to macroscopic physics (\emph{e.g.} astrophysics, cosmology) in certain regimes, is in terms of perfect fluids. Following Ref.~\cite{Brown:1992kc}, the matter action for a perfect and isentropic fluid is given by 
\be\label{eq:ActionMatter}
S_{\rm matter}=-\int\de^4 x\sqrt{-g} \left[\rho\left(\frac{|J|}{\sqrt{-g}}\right)+J^{\mu}(\pa_\mu\chi+\beta_A\pa_\mu\alpha^A)\right]~.
\ee
$J^{\mu}$ represents the densitized particle number flux (with $|J|\equiv\sqrt{-J^{\mu}J_{\mu}}$~), which can be written as
\be\label{eq:RelationFluxVelocity}
J^{\mu}=n\sqrt{-g}\,U^{\mu}~,
\ee
where $n$ is the particle number density and $U^{\mu}$ the four-velocity of the fluid. Using Eq.~(\ref{eq:RelationFluxVelocity}) the particle number density can be computed as
\be
n=\frac{|J|}{\sqrt{-g}}~.
\ee
$\chi$ is a Lagrange multiplier enforcing particle number conservation. Additional constraints can be imposed. In fact, interpreting $\alpha^A$ ($A=1,2,3$) as Lagrangian coordinates for the fluid, the Lagrange multipliers $\beta_A$ impose the condition that the fluid flows along lines of constant $\alpha^A$. The stress-energy tensor obtained from the action Eq.~(\ref{eq:ActionMatter}) takes the form
\be\label{eq:StressEnergy}
T_{\mu\nu}=-\frac{2}{\sqrt{-g}}\frac{\delta S_{\rm matter}}{\delta g^{\mu\nu}}= (\rho+p)\, U_\mu U_\nu + p\, g_{\mu\nu}~,
\ee
having defined the pressure as (see \cite{Brown:1992kc,Misner:1974qy})
\be
p=n\frac{\partial\rho}{\partial n}-\rho~.
\ee

The dynamics of the fluid is obtained by looking at the stationary points of the action (\ref{eq:ActionMatter}). In particular, diffeomorphism invariance implies that the stress-energy tensor is covariantly conserved
\be\label{eq:ConservationStressEnergy}
D^\mu T_{\mu\nu}=0.  
\ee
Notice that the local conservation law Eq.~(\ref{eq:ConservationStressEnergy}) is formulated in terms of the Levi-Civita connection. We remark that, as it is well-known, the argument above leading to Eq.~(\ref{eq:ConservationStressEnergy}) applies to all matter fields (including elementary ones, considered in the previous section) as long as interactions with other species are negligible. The Higgs field represents an exception, since it has direct couplings to the Weyl vector.

Different regimes have to be considered for the dynamics of matter, depending on the energy scale. Above the scale of EW symmetry breaking (and regardless of the fact that conformal symmetry is broken or unbroken), all particles are massless and can be described as a perfect radiation fluid $\rho_{\rm rad}(n)\propto n^{4/3}$. At lower scales and after the spontaneous breakdown of EW symmetry, photons and neutrinos remain massless, while baryonic matter\footnote{As in the ordinary usage of the word by cosmologists, \emph{i.e.} including leptons and actual baryons.} is characterized by $\rho_{\rm bar}(n)\propto n$. As far as the dynamics of matter fields alone\footnote{Again, with the exception of the Higgs field.} is concerned, there is no difference with the corresponding equations obtained in GR. Interactions with $B_\mu$ and $\phi$ can only be mediated by the gravitational field $g_{\mu\nu}$ or the Higgs field $H$. As it is well known, the dynamics of a small test body can be obtained from the conservation law Eq.~(\ref{eq:ConservationStressEnergy}) \cite{Geroch:1975uq}. This is readily seen for dust ($p=0$), in which case the worldline of each dust particle is a geodesic of the Levi-Civita connection, \emph{i.e.} the four-velocity satisfies the equation
\be
U^\mu D_\mu U^\nu=0~.
\ee

Geodesic motion of test bodies is a consequence of the coupled dynamics of the gravitational field and matter \cite{Geroch:1975uq}, not an independent physical principle. Hence, the connection that is used to define the parallel transport of \emph{physical} objects is \emph{not an independent prescription} fixed at the outset, but it is instead a consequence of the dynamics. Although this is a well-known result in General Relativity (see Ref.~\cite{Geroch:1975uq}), to the best of the authors' knowledge it has not been stressed previously in a non-Riemannian framework. In our case, the dynamics follows from an action principle which we built using local conformal invariance as an additional guiding principle. The Weyl connection is used as a tool to implement this principle in a natural way in the gravitational sector. It turns out that local conformal invariance in the sector of gauge bosons and spin-$1/2$ fermions does not require using a non-metric connection. The standard minimal coupling to the gravitational field is enough to ensure that conformal invariance holds as a local symmetry.

We would like to stress at this point that, although our approach is based on Weyl geometry as a framework for a dynamical theory of gravity, it differs from Weyl's original formulation in certain important respects. The main objection against Weyl geometry as a framework for gravitational physics is based on a criticism moved by Einstein against Weyl's original proposal. 
 Einstein's argument is the following. If a vector is parallel transported along a closed path, with parallel transport defined by the Weyl connection $\nabla_\mu$ instead of the Levi-Civita connection $D_\mu$, the norm of the vector changes as a result. This would have obvious physical consequences. In fact, considering any two paths in spacetime having the same starting and end points, rod's lengths and clock's rates would depend on their histories\footnote{The same argument would also apply for parallel transport given by other non-metric connections $\nabla_\lambda g_{\mu\nu}=Q_{\lambda\mu\nu}$ with non-vanishing Weyl vector, defined as the trace of the non-metricity $B_\mu=\frac{1}{4}Q^{\phantom{a}\phantom{a}\lambda}_{\mu\lambda}$.}. This is known as the \emph{second clock effect}. Any theory leading to such effects is clearly non-physical\footnote{The Aharonov-Bohm effect is an analogue of this effect which is instead physical. In that case though, the gauging is not done in physical space, as in Weyl's original proposal, but in the internal space given by the phase of the wave-function.}.

It is worth stressing that this is an argument against the use of the Weyl connection as the one defining parallel transport of \emph{physical} objects, such as rods and clocks. This is clearly not the case in our model. In fact, the dynamics of all elementary matter fields (with the important exception of the Higgs) only involves the Levi-Civita connection $D_\mu$. Hence, it does not entail any direct couplings to the new fields in the gravitational sector. Classical test particles move along geodesics defined by $D_\mu$, as in GR.
\section{An alternative proposal}\label{Section:Alternative}
In this section we suggest an alternative possibility for the dynamics of matter in the extended geometric framework of Weyl geometry. The reader must be aware that this proposal is entirely different in spirit from the one discussed in Sections~\ref{Sec:CouplingSM},~\ref{Sec:Fluids}. In fact, we will put aside for the time being the problem of finding a conformal invariant extension of the SM, and only focus on some classical aspects of the extended geometric framework. In particular, we will consider a different model to describe the motion of matter as classical test bodies. We assume a \emph{phenomenological} point of view and the existence of a conformal symmetry breaking mechanism from which mass scales originate. A specific coupling of classical test bodies with the Weyl vector is assumed, which is consistent with conformal symmetry in the unbroken phase.

We assume that the dynamics of a test particle is given by the following action
\begin{equation}\label{Eq:ActionTestParticle}
S_{\rm \scriptscriptstyle TP}=\frac{1}{2}\int\mbox{d} t\; \left[e^{-1}\dot{x}^\mu\dot{x}_\mu-m^2 e\right]-q\int\mbox{d} t\;B_\mu\dot{x}^\mu~.
\end{equation}
In Eq.~(\ref{Eq:ActionTestParticle}), $t$ is an arbitrary parameter on the world-line and the \emph{einbein} $e$ is a Lagrange multiplier. The second term represents the interaction of the test particle with the Weyl vector (with coupling $q$), which forms part of the extended gravitational background. In the conformally invariant phase, all dimensionful parameters must vanish. Hence, one has $m^2=0$ for all particles. The action (\ref{Eq:ActionTestParticle}) is then conformally invariant, with the metric and the Weyl vector transforming as in Eqs.~(\ref{eq:LocalConformalMetric}),~(\ref{eq:TransformationLawWeylVector}) and the einbein transforming as
\be
e\to\tilde{e}= \Omega^2 e~.
\ee
Variation of the action w.r.t. $e$ and $x^{\mu}$ in the massless case yields
\begin{align}
&\dot{x}^\mu\dot{x}_\mu=0~,\\
&\ddot{x}^\mu+\hspace{-5pt}\phantom{o}^{\rm \scriptscriptstyle LC}\Gamma_{\nu\kappa}^\mu\dot{x}^\nu\dot{x}^\kappa-eq H^{\mu}_{~\nu}\dot{x}^\nu=\left(\frac{\mbox{d}}{\mbox{d} t}\log e\right)\dot{x}^\mu~\label{eq:eom_massless}.
\end{align}
We can partially fix the world-line parametrization by requiring that the particle follows an affinely parametrized geodesic in the case $q=0$, which implies $\dot{e}=0$. We will denote the constant value of the einbein by $\hspace{-6pt}\ph^o e$. Making use of this additional assumption, Eq.~(\ref{eq:eom_massless}) thus reads
\be\label{eq:eom_massless:simplified}
\ddot{x}^\mu+\hspace{-5pt}\phantom{o}^{\rm \scriptscriptstyle LC}\Gamma_{\nu\kappa}^\mu\dot{x}^\nu\dot{x}^\kappa-\hspace{-6pt}\ph^o e q H^{\mu}_{~\nu}\dot{x}^\nu=0~.
\ee
Note that the coupling with the field strength in Eq.~(\ref{eq:eom_massless:simplified}) is entirely arbitrary. In fact, it depends on the time parametrization or, equivalently, on the choice of conformal frame. This freedom is essentially related to the fact that null curves are by definition invariant under conformal transformations, and to the absence of a basic time scale. We will come back to this issue later on.

In the broken-symmetry phase (\emph{i.e.} after conformal and EW SSB), mass scales are allowed. In this case, the dynamics is given by the action (\ref{Eq:ActionTestParticle}) with $m^2\neq0$. However, this is no longer conformally invariant. Solving the equations of motion for the einbein, the action (\ref{Eq:ActionTestParticle}) reduces to
\be\label{Eq:ActionTestParticle1}
\ph^{\scriptscriptstyle 1}S_{\rm \scriptscriptstyle TP}=-m\int\mbox{d} t\; \sqrt{-\dot{x}^\mu\dot{x}_\mu}-q\int\mbox{d} t\;B_\mu\dot{x}^\mu~.
\ee
Extremizing the action (\ref{Eq:ActionTestParticle1}) we obtain the equation of motion
\begin{equation}\label{Eq:EOMmassive}
\ddot{x}^\mu+\hspace{-5pt}\phantom{o}^{\rm \scriptscriptstyle LC}\Gamma_{\nu\kappa}^\mu\dot{x}^\nu\dot{x}^\kappa-\frac{q}{m} H^{\mu}_{~\nu}\dot{x}^\nu=0
\end{equation}
Note that in the massive case affine parametrization is automatically enforced when $q=0$. By comparing the equations of motion (\ref{Eq:EOMmassive}) and (\ref{eq:eom_massless:simplified}), we observe that there is in general a discontinuity in the coupling of the particle's velocity to the field strength $H^{\mu\nu}$ in the limit $m^2\to0$. In fact, for a fixed value of $q$ (\emph{i.e.} independent on $m$), the coefficient of $H^{\mu}_{~\nu}\dot{x}^\nu$ in Eq.~(\ref{Eq:EOMmassive}) diverges in the massless limit, whereas it is entirely arbitrary in the strictly massless case (Eq.~(\ref{eq:eom_massless:simplified})). However, if we assume that there is a continuous phase transition that gives rise to all mass scales, we can match the dynamics of the particle in the two phases by promoting $q$ to a function of the mass and requiring $q\propto m$. In principle, the proportionality constant can depend on the the internal constitution of the test body. However, if we assume that it is universal, the motion of test bodies is the same as in the scalar-tensor-vector theory (MOG) of Ref.~\cite{Moffat:2005si} provided that $q/m=\kappa_g=\sqrt{\alpha G}$ (for the definition of the parameters\footnote{The parameter $\kappa_g$ used in Ref.~\cite{Moffat:2005si} should not be confused with the parameter $\kappa$ used in the rest of this paper, \emph{e.g.} in Eq.~(\ref{eq:Higgs-Dilaton}).} $\alpha$ and $\kappa_g$ see Ref.~\cite{Moffat:2005si}).

Some remarks are in order:
\begin{enumerate}[label=$\roman*$)]
\item This approach, as the one discussed previously in Sections~\ref{Sec:CouplingSM},~\ref{Sec:Fluids}, is also immune from the second clock problem, but for a different reason. In fact, the motion of test particles in this case does not follow geodesics of the Levi-Civita connection, but is also influenced by the Weyl vector. However, only its field strength $H_{\mu\nu}$ appears in the equations of motion (\ref{eq:eom_massless}),~(\ref{eq:eom_massless:simplified}),~(\ref{Eq:EOMmassive}). Hence, the rate of a clock does not change by going around a closed path.
\item Despite the formal analogy of the action (\ref{Eq:ActionTestParticle1}) with that of a charged particle in classical electrodynamics, the Weyl vector $B_\mu$ should not be identified with the electromagnetic field potential $A_\mu$. In fact, although their transformation properties are similar (under local conformal transformations vs. gauge transformations), the other fields (\emph{e.g.} the metric tensor) do not behave in the same way under a conformal or an internal $\mbox{U}(1)$ transformation.
\item The approach followed in this Section is a phenomenological one, which assumes a strictly macroscopic point of view. No attempt is made to connect it to an underlying field theory which is compatible with conformal invariance. In fact, as shown in Sections~\ref{Sec:CouplingSM},~\ref{Sec:Fluids}, minimal conformal invariant extensions of the SM and GR do not lead to similar dynamics for test bodies. Rather, they imply that test bodies follow geodesics of the Levi-Civita connection as in GR.
\item It is not clear yet how a universal coupling to $B_\mu$ with coupling constant $q\propto m$ may arise from the point of view of quantum field theory, in a way which is at the same time compatible with the conformal symmetry breaking scenario outlined in the previous sections, and with the Higgs mechanism.
\end{enumerate}

\section{Conclusion}\label{Conclusions}
We considered an extension of GR and SM with local conformal invariance. The purpose is to provide a new framework for the study of conformal symmetry and its relation to fundamental physics at high energy scales. This is achieved by considering a generalization of Riemannian geometry, first introduced by Weyl and later proposed by Smolin~\cite{Smolin:1979uz}. The affine connection is no longer given by the Levi-Civita connection, as only the conformal structure of the metric is preserved by parallel transport. This leads to the introduction of a gauge vector $B_{\mu}$ in the gravitational sector: the Weyl vector. A scalar field $\phi$ is also needed in order to build a conformally invariant action functional. The framework is that of a classical effective field theory of gravity. The interpretation of our model is similar to that of quantum field theory in curved spacetime. SM fields can be quantized as usual, with $g_{\mu\nu}$, $B_{\mu}$ and $\phi$ representing \emph{classical} background fields\footnote{This is clearly the case for the metric $g_{\mu\nu}$ and the Weyl vector $B_{\mu}$ since they define the classical geometric structure of spacetime, see Eq.~(\ref{eq:DefWeylConnection}). In fact, either they are both classical or both quantum. The status of the field $\phi$ is a more subtle issue and both cases are possible \emph{a priori}. Only a careful analysis of the implications of the two possibilities can determine which one is correct.}.

Our model is a generalization of previous works in the scientific literature on conformal symmetry in gravity theories \cite{Bars:2013yba,Kallosh:2013hoa}, which can be recovered as a particular case of our model. The main difference in our approach is due to the introduction of a new geometric degree of freedom, represented by the Weyl vector field entering the definition of the Weyl connection. Suitable choices of some parameters of the theory lead to the decoupling of the Weyl vector from the Higgs and the scalar fields. Although, in the general case its dynamics cannot be neglected. After gauge fixing the conformal symmetry (which can be interpreted as a spontaneous symmetry breaking) and EW symmetry breaking, the Weyl vector acquires a mass and the scalar is completely decoupled from the theory. The relevance of the scalar for low energy physics lies in the fact that, through gauge fixing, it leads to the introduction of a physical scale in a theory which is scale-free at the outset. All dimensionful parameters of the SM and gravity can be expressed in terms of it and of the dimensionless parameters of the theory.

Einstein's criticism to Weyl's original proposal is addressed in our model, which is not affected by the \emph{second clock effect}. In fact, we showed in Sections~\ref{Sec:CouplingSM},~\ref{Sec:Fluids} that the affine connection that defines parallel transport of physical obejcts, such as \emph{e.g.} clocks and rods, is the Levi-Civita connection. Test particles move along Levi-Civita geodesics as in GR. We remark that this is not prescribed at the outset. It is instead a consequence of the dynamics, which has been formulated using conformal invariance as a guiding principle. The Weyl connection \emph{does} play a role in determining the gravitational sector of the theory, although it does not determine the motion of test particels\footnote{It is remarkable that essentially the same observation was made by Weyl in a reply to Einstein's comment to his original paper. We quote from the English translation contained in Ref.~\cite{ORaifeartaigh:1997dvq}: \emph{``It is to be observed that the mathematical ideal of vector-transfer {\rm (Authors' Note: \emph{i.e.}, parallel transport)}, on which the construction of the geometry is based, has nothing to do with the real situation regarding the movement of a clock, which is determined by the equations of motion''}.}. Furthermore, the introduction of the $B_\mu$ field is necessary in order to build a conformally invariant action functional for scalar fields in four dimensions, but has no (direct) effects on radiation and baryonic matter.
SM fields do not couple to the new fields in the gravitational sector, with the exception of the Higgs. Their interactions with $\phi$ and $B_\mu$ can only be mediated by gravity or the Higgs field.

\section{Discussion and Outlook}\label{Sec:Discussion}
We would like to stress that the present model does not necessarily offer a resolution of the naturalness (or hierarchy) problem. In fact, such problem is now translated in the fine-tuning of its dimensionless parameters. Namely, the hierarchy of the Planck versus EW scale leads to $\frac{v^2}{M_{Pl}^2}=\frac{\xi_\phi}{2\kappa^2}+\xi_H\sim 10^{-34}$. Nevertheless, classical conformal invariance of the extended SM sector is important as a guideline for model building, since it restricts the class of allowed couplings to those having dimensionless coupling constants \cite{Heikinheimo:2013fta}. Furthermore, the possibility of addressing the hierarchy problem in conformally invariant extensions of SM has been considered in \emph{e.g.} Ref.~\cite{Meissner:2006zh} and in earlier works Refs.~\cite{Buchmuller:1988cj,Buchmuller:1990pz}. In the models considered in those works, the EW and the Planck scales are determined by non-trivial minima of the one-loop effective potential in the Higgs-dilaton sector\footnote{The mechanism is a generalization of the one originally proposed by Coleman and Weinberg in Ref.~\cite{Coleman:1973jx}.}. It will be the subject of future work to study whether a similar mechanism could be implemented consistently within our framework. In fact, whereas it is clear that the Weyl vector cannot be quantized without also quantizing the metric, one may speculate that the scalar field $\phi$ should be treated on the same footing of matter fields and be regarded as quantum. Hence, similar analysis as in the works cited above should be carried out to check the viability of such working hypothesis. In the affirmative case, it would be possible to address the important point concerning the exact value of the scale $\phi_0$, which we regarded as a free parameter in this work\footnote{Similarly, the scale of ``conformal symmetry breaking'' (gauge fixing) $\phi_0$ is a free parameter in all models with classical conformal invariance see \emph{e.g.} Refs.~\cite{Bars:2013yba,Kallosh:2013hoa}.}.

In future work we will explore the physical consequences of our model for cosmology and astrophysics. In particular, it would be interesting to study whether $B_\mu$ could represent a valid dark matter candidate, as it was first hinted by \cite{Cheng:1988zx}. If this was the case, it would have a substantially different interpretation from standard dark matter. In fact, the Weyl vector should not be regarded, strictly speaking, as matter but as a property of the spacetime geometry. Important viability checks for the model require the determination of constraints on the couplings of $\phi$ and $B_\mu$ to the Higgs that may come from collider physics. The Weyl vector $B_\mu$ is a classical background field; hence, it can only contribute external lines to the diagrams describing known processes. This is true also for the scalar $\phi$, if this is to be regarded as classical. If, on the other hand, $\phi$ is treated as a quantum field, there will be a new scalar entering loop diagrams. In this case, it is crucial to determine which values of the coupling constants (such as \emph{e.g.} $\xi_\phi$, $\xi_H$) in the bare action are such that renormalizability of SM is not spoiled (see \emph{e.g.} the analysis in Ref.~\cite{Coriano:2012nm}). Addressing this question may also help to shed some light on the ``naturalness'' of the particular choice of parameters\footnote{Commonly known as conformal couplings, since in the standard framework of Riemannian geometry these are the unique values that make the kinetic terms of $\phi$ and $H$ conformally invariant.} $\xi_H=\frac{\xi_\phi}{\omega}=-\frac{1}{12}$ within the broader framework of Weyl geometry. The phenomenology of the model should be studied in detail both in the case where $\phi$ is quantized and when it represents instead a classical background. Detailed studies of the consequences for gravitational experiments are also in order and will be the subject of future work. In particular, we plan to explore in a future work the possible observable consequences of the enriched gravity sector and its implications for astrophysics and cosmology.

It is also worth studying the possible relations between our model and modified gravity theories such as the scalar-tensor-vector theory (MOG) considered in Ref.~\cite{Moffat:2005si}. A preliminary study of the possibility of such a connection was carried out in Section~\ref{Section:Alternative}. This was done by considering a purely phenomenological model for the dynamics of test bodies, which could represent an alternative scenario to the theoretical model analyzed in Sections~\ref{Sec:CouplingSM},~\ref{Sec:Fluids}.
\begin{acknowledgement}
The research was supported in part by Perimeter Institute for Theoretical Physics. Research at Perimeter Institute is supported by the
Government of Canada through the Department of Innovation, Science and Economic Development and by the Province of Ontario through
the Ministry of Research and Innovation. The work of MS is also partially supported by STFC (UK) under the research grant
ST/L000326/1. MdC and MS would like to thank Simon Salamon for useful discussions and Perimeter Institute for the hospitality while this work was completed. MdC is also grateful to Roberto Oliveri and J\'er\'emie Quevillon for their constructive comments on the draft. JWM would like to thank Philip Mannheim for useful discussions. The authors would like to thank Subhash Rajpoot for drawing our attention to his work with Hitoshi Nishino. The authors wish to acknowledge helpful feedback received from the anonymous referees, that prompted us to clarify some crucial physical aspects of the model.
\end{acknowledgement}
\appendix
\section*{Appendix}\label{Appendix}
In this Appendix we wish to add more details showing the motivation for considering the action (\ref{Eq:ActionDirac2}). Furthermore, we will prove that the two Dirac Lagrangians in Eqs.~(\ref{eq:ActionDirac}) and (\ref{Eq:ActionDirac2}) are the same, as already pointed out in Ref.~\cite{Cheng:1988zx}.

The reason to look at the action (\ref{Eq:ActionDirac2}) in first place, is to have an action functional which is manifestly conformally invariant. In fact, given a field $F$ with conformal weight $w$, \emph{i.e.} transforming under a local conformal transformation as
\be\label{eq:ConformalWeight}
F\rightarrow\tilde{F}=\Omega^{w}F~.
\ee
For scalar fields and half-spin fermions the conformal weight is the opposite of their canonical mass-dimension, \emph{i.e.} $w=-1,-3/2$, respectively\footnote{One must be aware that this statement cannot be generalized to fields with arbitrary canonical mass-dimension. In fact, for a gauge vector field, one must have $w=0$, see the discussion in Section and Ref.~\cite{Wald:1984rg}.} . Therefore, we can construct a ``gauge covariant derivative'' for the conformal symmetry as
\be\label{eq:CovariantDerivative}
\mathcal{D}_\mu F=\pa_\mu F -\frac{w}{2}B_\mu F~.
\ee
It is straightforward to check, using Eqs.~(\ref{eq:TransformationLawWeylVector}),~(\ref{eq:ConformalWeight}),~(\ref{eq:CovariantDerivative}), that
\be
\mathcal{D}_\mu F\rightarrow\tilde{\mathcal{D}}_\mu \tilde{F}=\Omega^{w}\,\mathcal{D}_\mu F~,
\ee
which justifies calling $\mathcal{D}_\mu$ a gauge covariant derivative. This is all we need to build the kinetic term of a scalar field in Eq.~(\ref{eq:PhiSector}), but it is not enough for fermions. In fact, the spin connection must appear explicitly in the action of a spinor. In the metric-compatible case, the spin connection is given by (see Ref.~\cite{Wald:1984rg})
\be\label{eq:LCSpinConnection}
\omega^{\rm\scriptscriptstyle LC}_{\mu\, ab}=e_a^{\;\nu}D_\mu e_{b\,\nu}~.
\ee
In order to be consistent with the principle of local conformal invariance, it is natural to replace this object with the one constructed out of the Weyl connection
\be\label{eq:WeylSpinConnection}
\omega^{\rm\scriptscriptstyle W}_{\mu\, ab}=e_a^{\;\nu}\nabla_\mu e_{b\,\nu}~.
\ee
Under a conformal transformation we have
\be
\omega^{\rm\scriptscriptstyle W}_{\mu\, ab}\rightarrow\tilde{\omega}^{\rm\scriptscriptstyle W}_{\mu\, ab}=\omega^{\rm\scriptscriptstyle W}_{\mu\, ab}+\left(\Omega^{-1}\pa_{\mu}\Omega\right)\eta_{ab}~.
\ee
Notice that in the case of Weyl geometry, the spin connection fails to be antisymmetric in the internal indices $\omega^{\rm\scriptscriptstyle W}_{\mu\, ab}\neq-\omega^{\rm\scriptscriptstyle W}_{\mu\, ba}$, as it is instead the case in Riemannian geometry.
However, since in the Dirac action (\ref{Eq:ActionDirac2}) $\omega^{\rm\scriptscriptstyle W}_{\mu\, ab}$ is contracted with the generator of Lorentz transformations in spinor space (which is $\propto [\gamma^a,\gamma^b]$), only the antisymmetric part gives a non-vanishing contribution\footnote{Which is also the reason why it does not make a difference whether we define $\omega^{\rm\scriptscriptstyle W}_{\mu\, ab}$ as equal to $e_a^{\;\nu}\nabla_\mu e_{b\,\nu}$ or $e_{a\nu}\nabla_\mu e_{b}^{\;\nu}$, although they are not the same in the non metric-compatible case.}. Hence, the third term in the bracket in the action (\ref{Eq:ActionDirac2}) is conformally invariant.

We will now proceed to show that the Lagrangian in Eq.~(\ref{Eq:ActionDirac2}) is equal to the Dirac Lagrangian in the metric case, which appears in Eq.~(\ref{eq:ActionDirac}). In order to do this, we expand the spin connection in Eq.~(\ref{eq:WeylSpinConnection}) in terms of its counterpart in the metric case, given by Eq.~(\ref{eq:LCSpinConnection}), plus terms involving the Weyl vector
\be
\omega^{\rm\scriptscriptstyle W}_{\mu\, ab}=\omega^{\rm\scriptscriptstyle LC}_{\mu\, ab}+e_{[a}^{\; \nu}e_{b]\mu}B_{\nu}+\frac{1}{2}\eta_{ab}B_{\mu}~.
\ee
Hence, we have for the contribution to the action (\ref{Eq:ActionDirac2}) coming from the last term in the round bracket
\be
\frac{1}{8}\gamma^c e_c^{\;\mu}[\gamma^a,\gamma^b]\,\omega^{\rm\scriptscriptstyle W}_{\mu\, ab}=\sum_{c\neq a}\frac{1}{4}\gamma_c\gamma^a\gamma^c e_{a}^{\; \nu}B_\nu
=-\frac{3}{4}\gamma^a e_{a}^{\; \nu}B^{\nu}~,
\ee
which cancels exactly the contribution due to the gauge-covariant coupling to $B_{\mu}$.
 \bibliographystyle{hieeetr}
 \bibliography{biblioweyl2}
\end{document}